\input{aipcheck}
\documentclass[
    ,final            
  ]
  {aipproc}

\layoutstyle{6x9}
\usepackage{subfigure}

\begin{document}
\title{Commissioning of the ATLAS \\ Liquid Argon Calorimeters}

\classification{29.40}
\keywords      {ATLAS, Liquid Argon, Calorimeter, Commissioning, Cosmic Muons}

\author{Mark S. Cooke \\  \small{(on behalf of the ATLAS Liquid Argon Calorimeter Group)}}{
  address={Columbia University, New York, NY 10027 USA}
}

\begin{abstract}
A selection of ATLAS liquid argon (LAr) calorimeter commissioning studies is presented.  It includes a coherent noise study, a measurement of the quality of the ionization pulse shape prediction, and energy and time reconstruction analyses with cosmic and single beam signals.
\end{abstract}

\maketitle


\section{Introduction}
ATLAS\cite{Aad2008} is a general purpose detector designed to study ${pp}$ collisions at $\sqrt{s} = 14$ TeV produced by the Large Hadron Collider (LHC) at CERN.  The detector utilizes liquid argon (LAr) calorimetry for the electromagnetic (EM) barrel and end-cap calorimeters, as well as for the hadronic end-cap (HEC) and forward calorimeters (FCAL).  

The first modules of the ATLAS LAr calorimeters were read out \textit{in situ} in March 2006.  Since then the full system has been brought online and regular calibration and cosmic muon runs have been taken and analyzed. Additionally, a sample of approximately 100 single-beam-on-collimator events were recorded during the initial LHC start-up in September 2008.  Results from several commissioning studies are presented.

\begin{figure}[]
\begin{minipage}[b]{0.5\linewidth} 
\centering
\includegraphics*[scale=0.35]{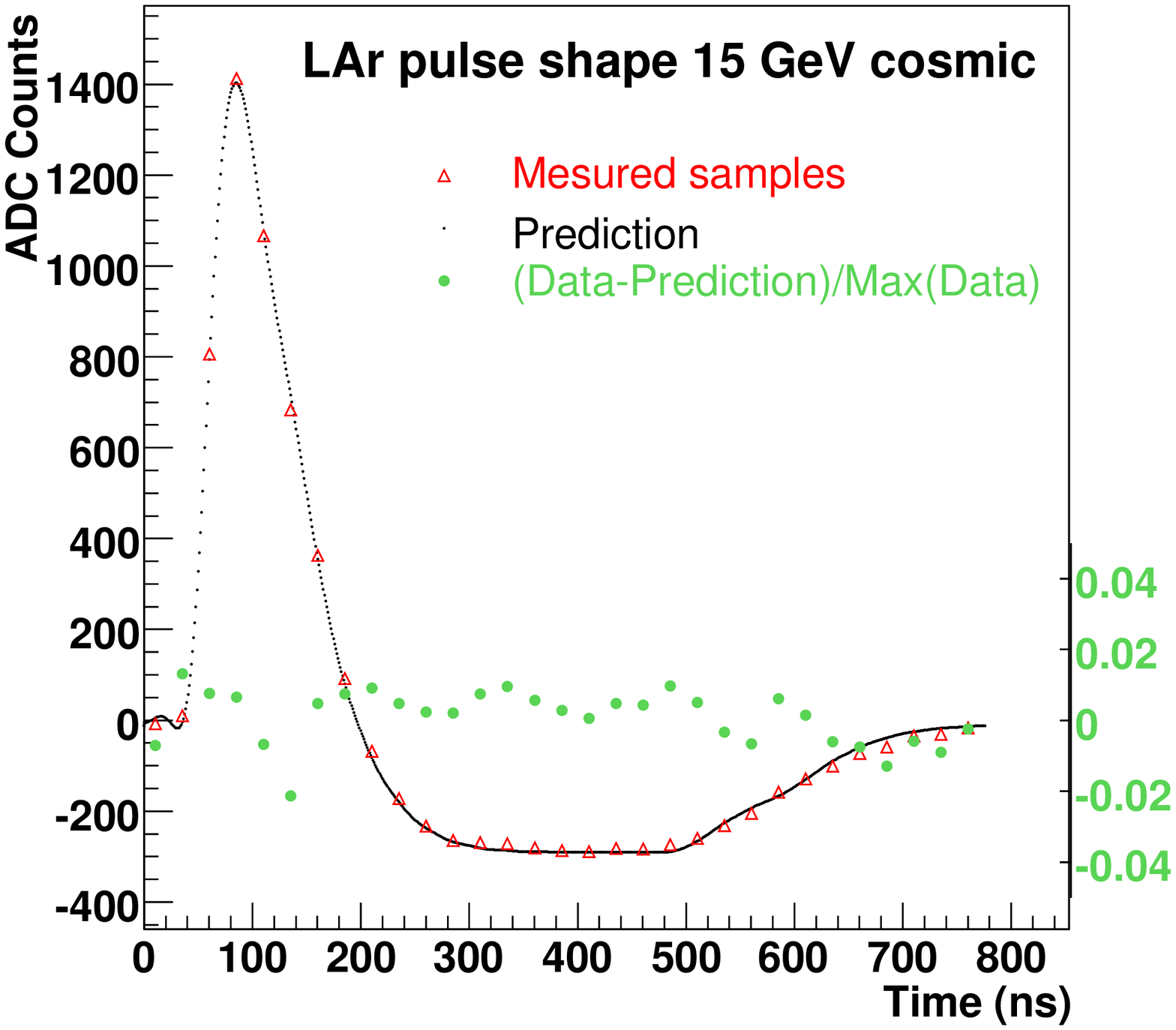}
(a)
\end{minipage}
\begin{minipage}[b]{0.5\linewidth}
\centering
\includegraphics*[scale=0.34]{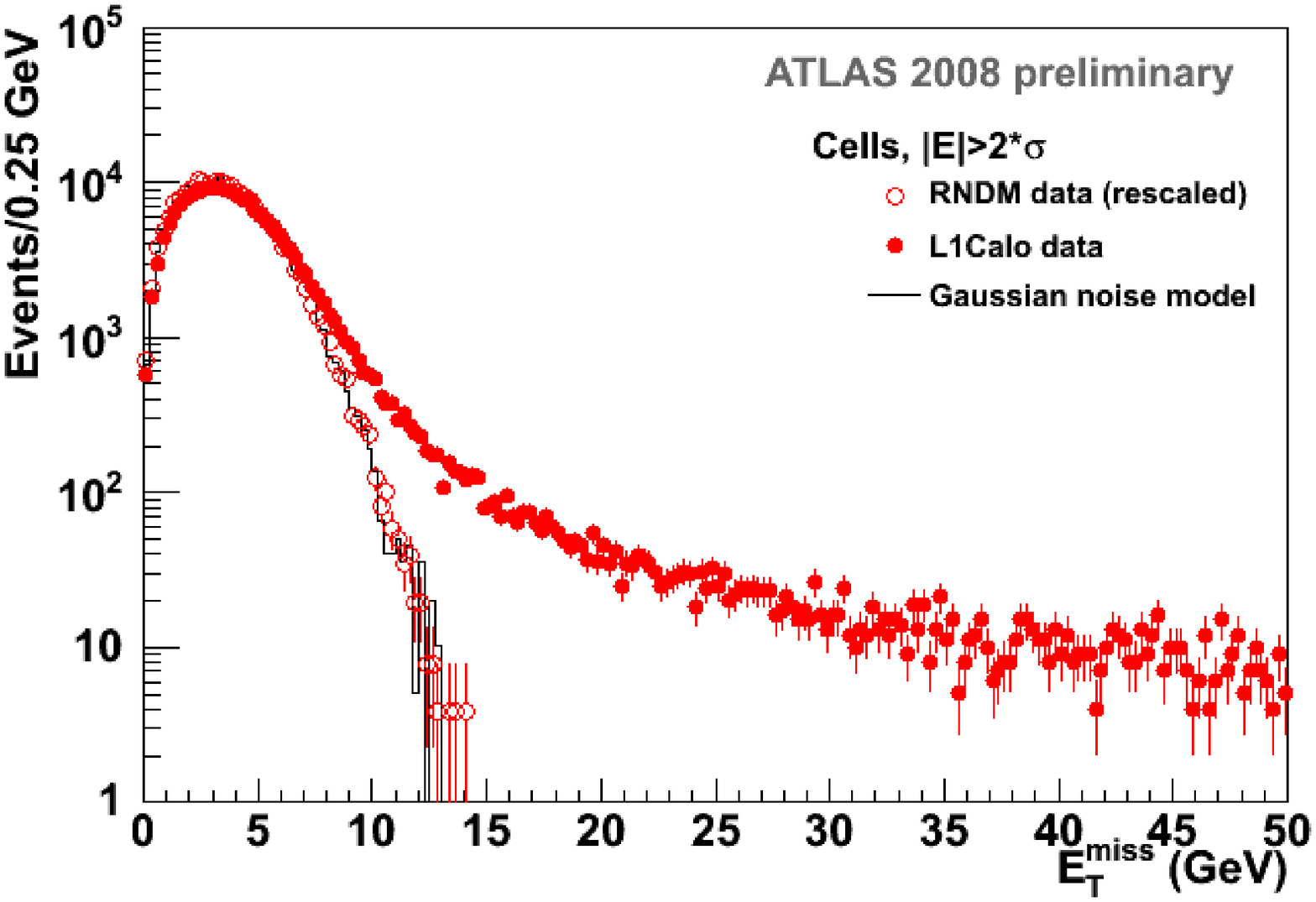}
(b)
\end{minipage}
\caption{(a) High amplitude shaped and digitized ionization pulse from a cosmic event in which the muon underwent catastrophic energy loss.  The pulse shape prediction is shown overlaid. The residuals are less than 2\% over the full length of the pulse. (b) The $E_{T}^{miss}$ distribution calculated from LAr cells.}
\end{figure}

\section{Signal Processing of the Ionization Pulse}

Absorber plates in the calorimeters initiate the particle shower development which ionizes the LAr medium.  High voltage is applied and drifts the ionized electrons creating a current which is read out. Front end boards\cite{Buch2008} (FEBs) located on the detector receive and process the ionization signals.  

The FEBs amplify and shape the signals in 3 gains, sample and analog store them during the level 1 (L1) trigger latency, gain select, digitize upon L1 accept, and transmit the samples (typically 5) to read out drivers off the detector.  Figure 1(a) shows a shaped and digitized waveform for a cosmic signal (taken in 32 sample read out mode).

\section{Commissioning Results}

The studies presented here address the topics of detector noise, calibration, and energy and time reconstruction with cosmic and single beam signals. \\

The noise of a LAr cell is dominated by the FEB pre-amp noise loaded by the capacitance of the cell.  The cell capacitance is different for EM, HEC, and FCAL calorimeters, and also varies with $\eta$ and longitudinal layer within a detector element.  The noise of each cell is occasionally measured and stored into a database. Using these noise values, the $E_{T}^{miss}$ distribution is predicted assuming incoherent noise between channels.  This is shown as the Gaussian noise model in Fig. 1(b).  The measured $E_{T}^{miss}$ in randomly triggered events is also shown and agrees well with the incoherent noise prediction.  A cosmic tail (checked by pulse shapes) is observed in L1Calo triggered events. \\

A calibration system \cite{Colas2008} is used to determine the ionization pulse shape prediction shown in Fig. 1(a).  An analysis \cite{Ruiz2008} of high amplitude cosmic muon signals in the EM calorimeter has demonstrated that the calibration procedure \cite{Collard2007} accurately models the response over the full $\eta$ coverage.  The quality of the pulse shape prediction is quantified by the formula
\begin{equation}
Q^{2}=\frac{1}{nDoF}\sum_{i=1}^{n_{samples}}\frac{(A_{i}^{data}-A_{i}^{pred})^{2}}{\sigma^{2}_{noise}+\sigma^{2}_{pred}},
\end{equation}
where $A_{i}^{data}$ are the measured samples and $A_{i}^{pred}$ the corresponding prediction.
The analysis was recently performed on single beam data and is shown in Fig. 2(a).  An important free parameter in the pulse shape prediction is the drift time.  The drift time was extracted from fits to a large number of cosmic pulses and shown to depend on $\eta$ as a result of $\sim100$ $\mu\mathrm{m}$ displacements of the electrodes. The measured drift time versus $\eta$ is shown in Fig. 2(b).  The data points are the means of Gaussian fits and the continuous line is the prediction from absorber thickness measurements.

\begin{figure}[]
\begin{minipage}[b]{0.5\linewidth} 
\centering
\includegraphics*[scale=0.27]{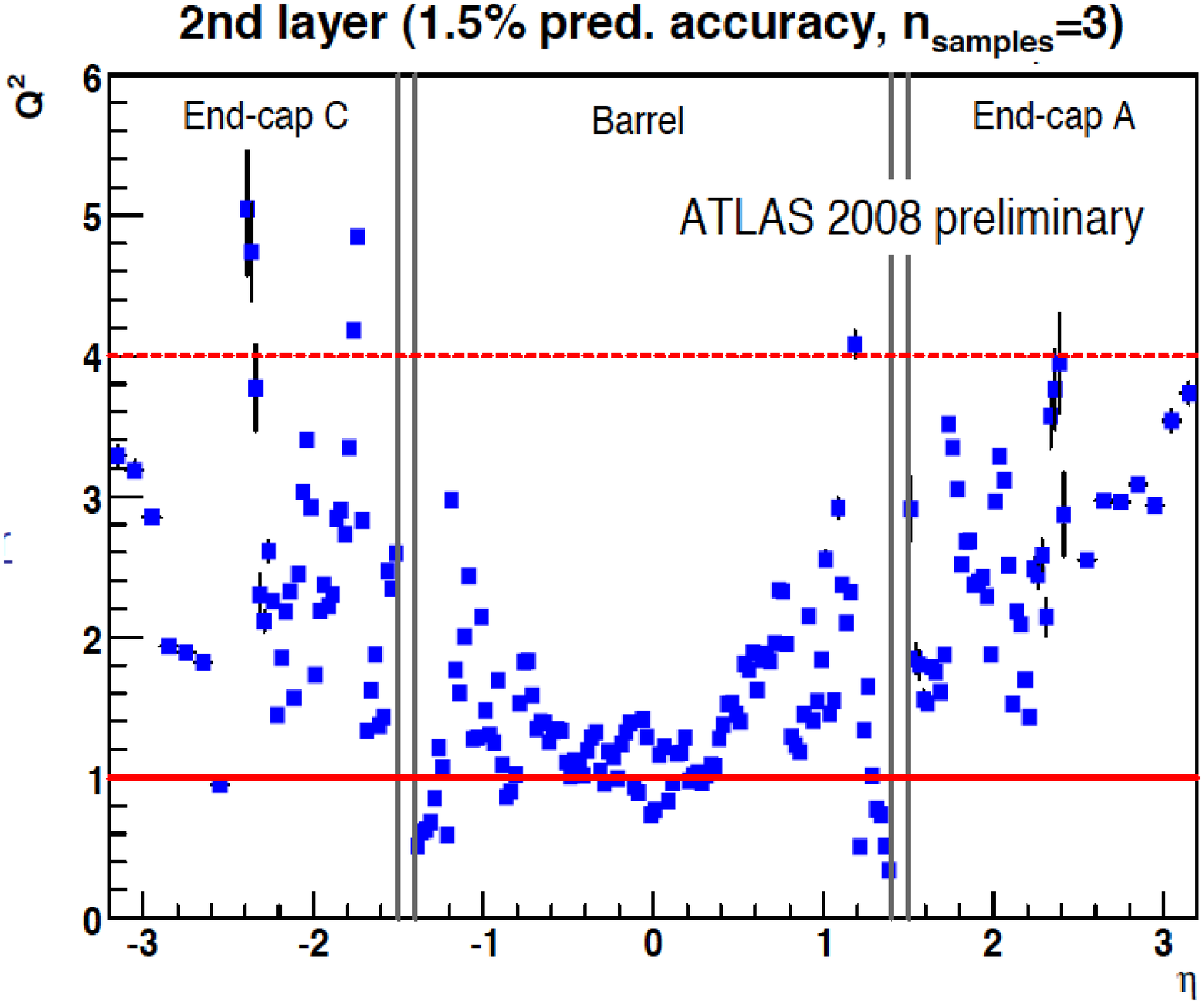}
(a)
\end{minipage}
\hspace{0.3cm}
\begin{minipage}[b]{0.5\linewidth}
\centering
\includegraphics*[scale=0.36]{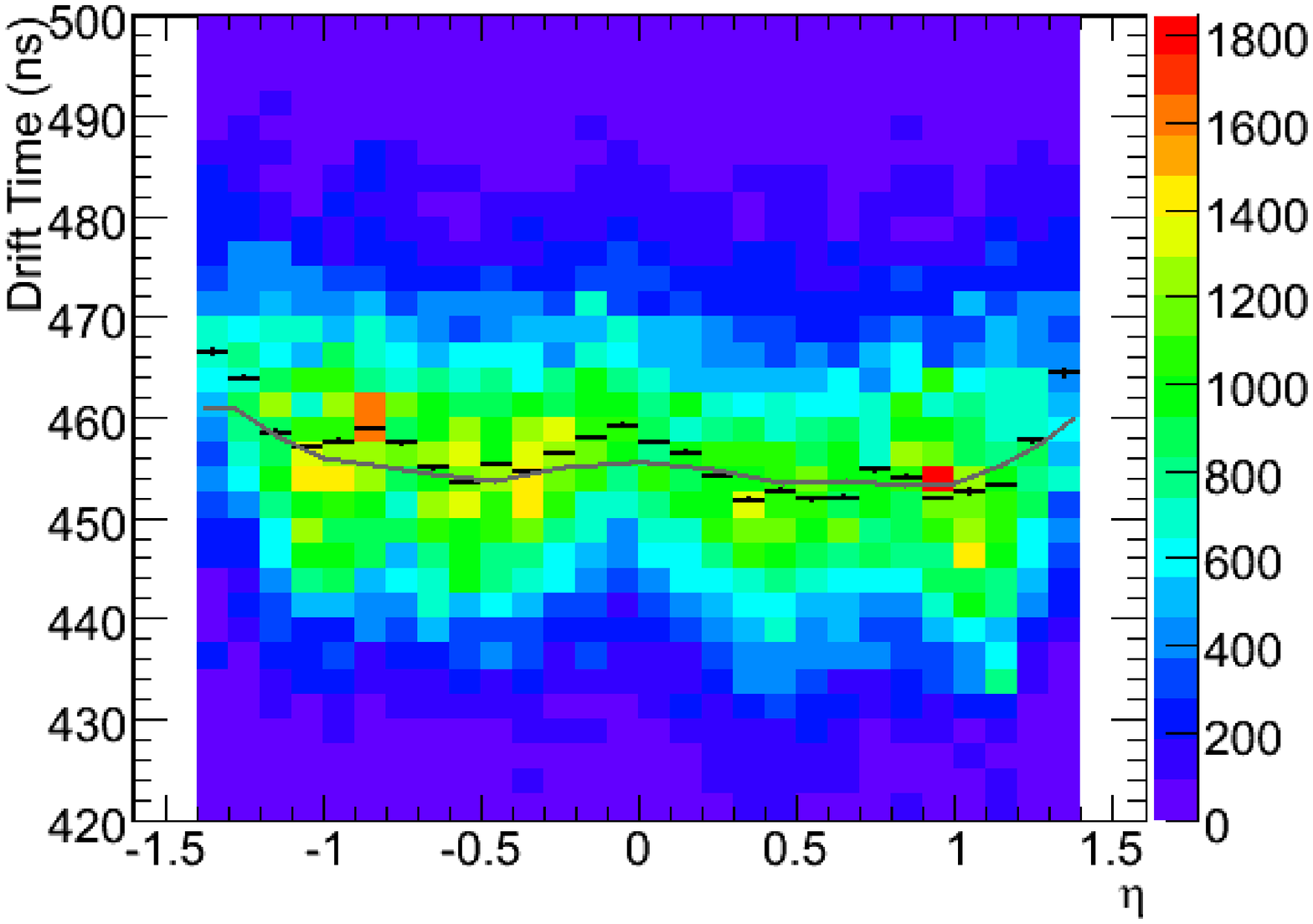}
(b)
\end{minipage}
\caption{(a) The quality of the physics pulse shape prediction, $Q^{2}$, as given by Equation (1) , as a function of $\eta$. (b) The measured drift time as a function of $\eta$.}
\end{figure}

Muons are minimum ionizing particles and typically leave small energy depositions in the LAr detectors.  The energy deposition follows a Landau distribution and the most probable value (MPV) scales linearly with the distance over which the energy is deposited.  Figure 3(a) shows the distribution of cluster energy in the range $0.3<|\eta|<0.4$ for two cluster algorithms\footnote{LArMuID is a variable size cluster better suited for normal LHC running, while the 3x3 cluster is less sensitive to out-of-cluster energy loss for less projective muons.  The bias is observed in Fig. 3(a) and 3(b).}\cite{Cooke2007}.  The distributions are fit to a Landau convoluted with a Gaussian.  Figure 4(b) shows the MPV as a function of $\eta$.  The variation with $\eta$ is a result of the non-uniformity of the cell depth.  The $\eta$ dependence and overall energy scale of the 3x3 method agree with the Monte Carlo to within 3\%. \\

Single-beam-on-collimator events in September 2008 deposited significant energy over large portions of the LAr detectors. Figure 4(a) shows the accumulated energy in these events per layer 2 EM cell.  The 8-fold pattern in $\phi$ is a result of the absorption of energy by the end-cap toroids between the collimator and calorimeters.  These events produced a large number of high amplitude pulses with a common reference time which proved useful for timing alignment studies.  Each cell time is corrected for the time of flight (TOF) from the collimator.  This time is compared with an expected time obtained from TOF from the origin and known delays in the readout.  Figure 4(b) shows the comparison\footnote{ Figure 4(b) is shown as a function of the FEB slot variable which is related to regions of $\eta$ and longitudinal layer.}
and the agreement is at the level of 2ns \cite{Guillemin2009}. Deviations between the measurement and the prediction can be corrected with a progammable delay on the FEB.\\

In conclusion, the commissioning results presented here reflect the readiness of the ATLAS LAr calorimeters for LHC collisions.

\begin{figure}[]
\begin{minipage}[b]{0.5\linewidth}
\centering
\includegraphics*[scale=0.38]{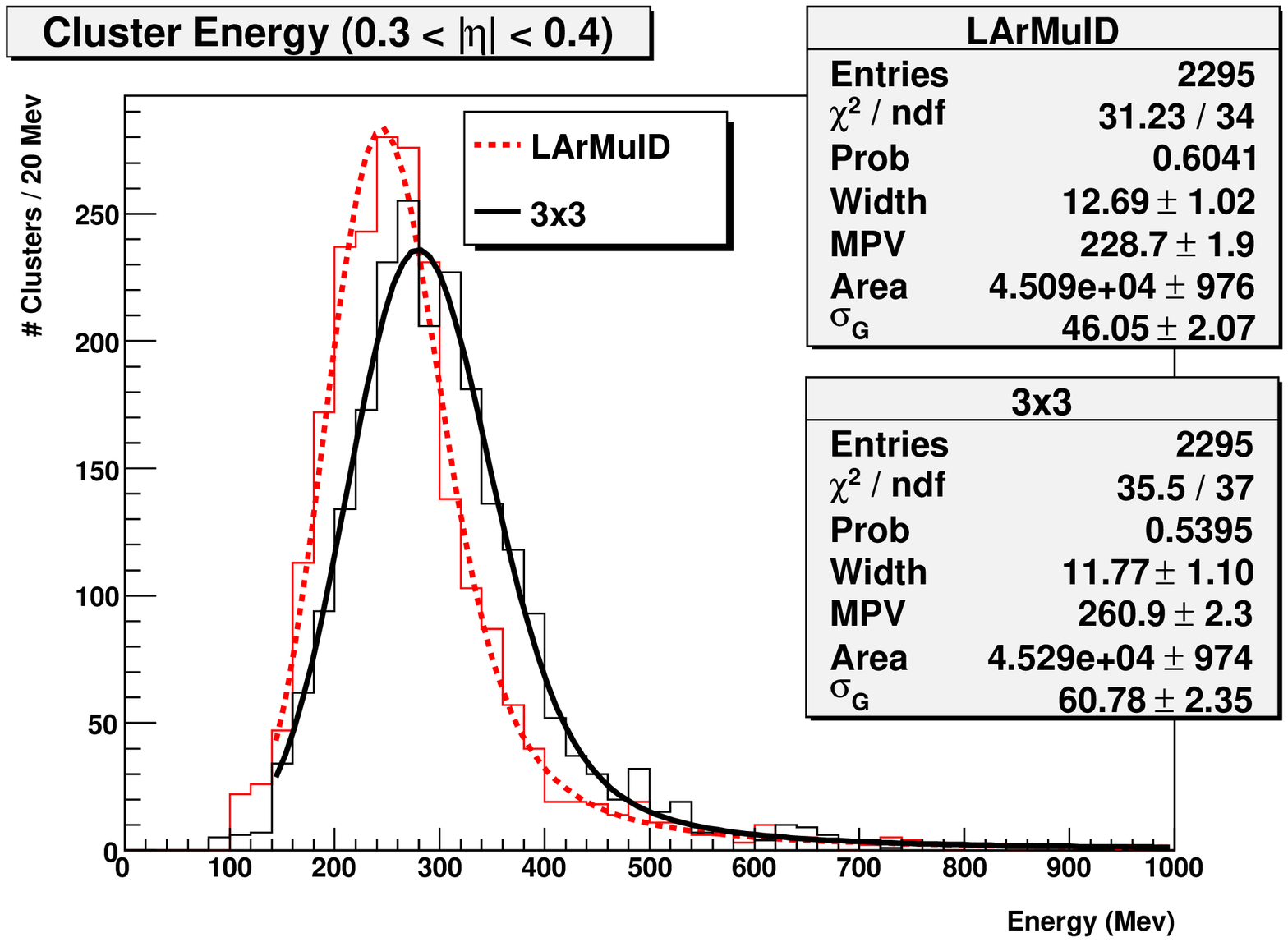}
(a)
\end{minipage}
\hspace{0.5cm}
\begin{minipage}[b]{0.5\linewidth}
\centering
\includegraphics*[scale=0.34]{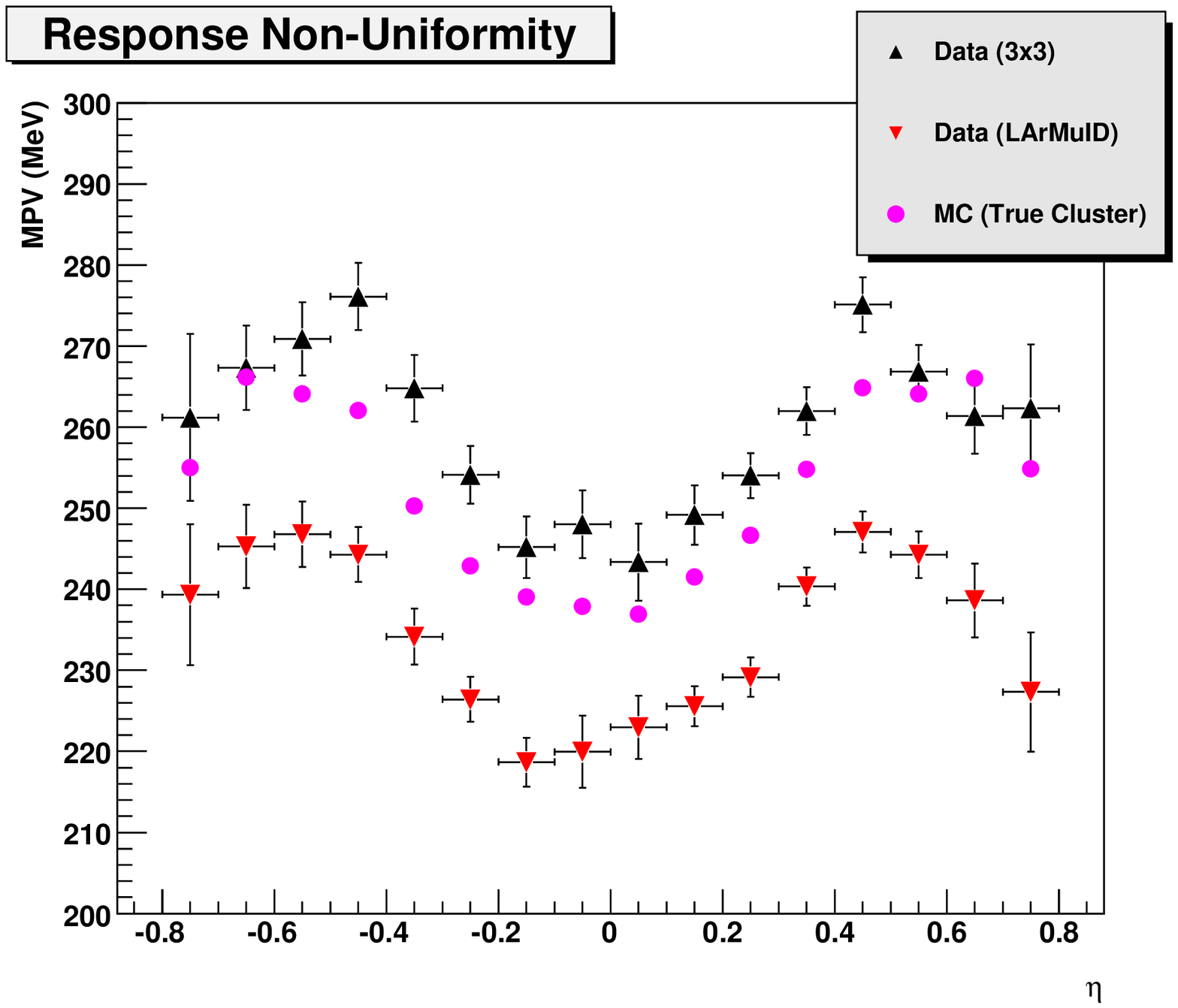}
(b)
\end{minipage}
\caption{(a) Cluster energies (2 cluster methods) in the range $0.3<|\eta|<0.4$ in semi-projective cosmic events. (b) The most probable value fit to the distributions as shown in (a) as a function of $\eta$.}
\end{figure}

\begin{figure}[]
\begin{minipage}[b]{0.5\linewidth}
\centering
\includegraphics*[scale=0.38]{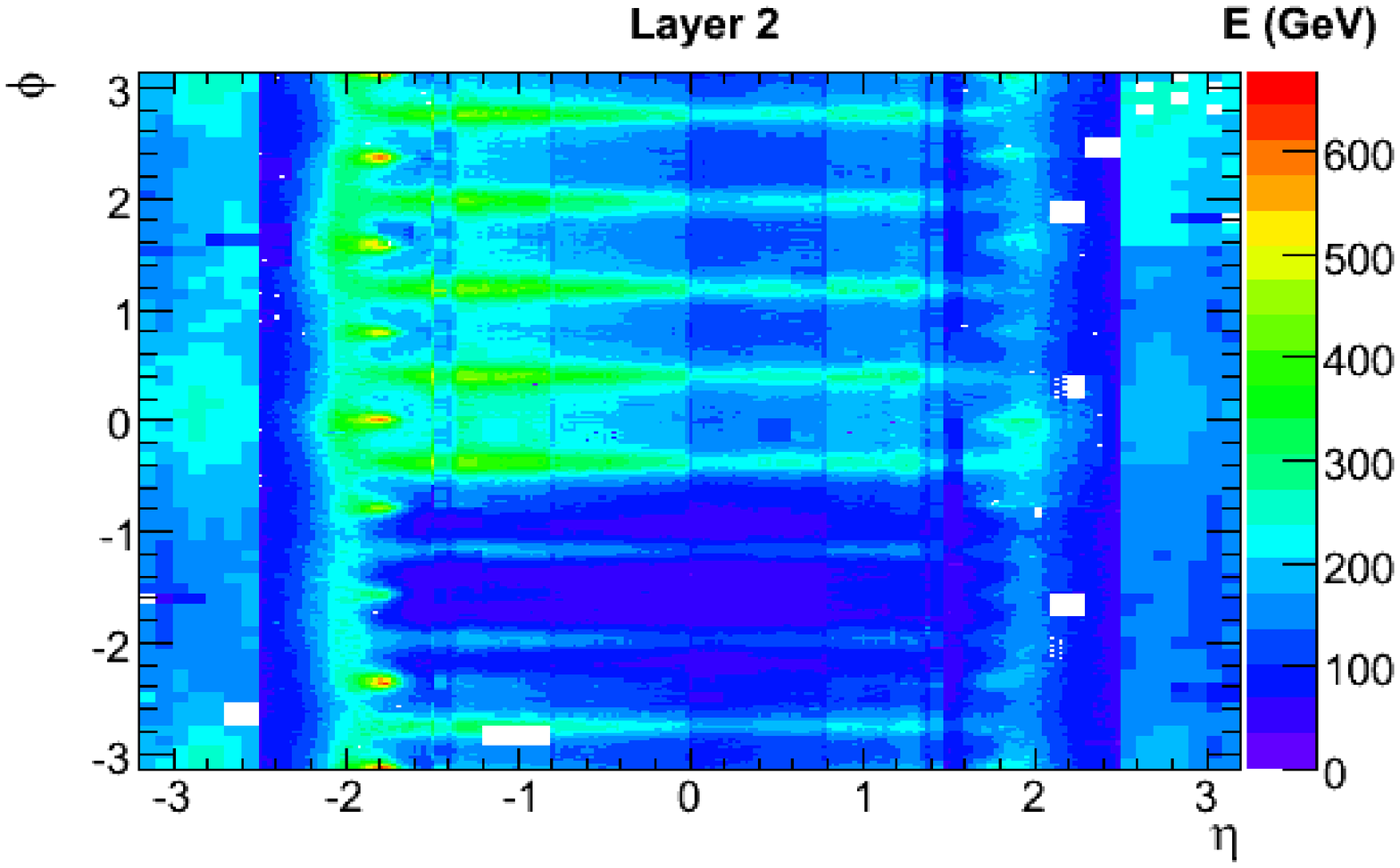}
(a)
\end{minipage}
\hspace{0.5cm}
\begin{minipage}[b]{0.5\linewidth}
\centering
\includegraphics*[scale=0.32]{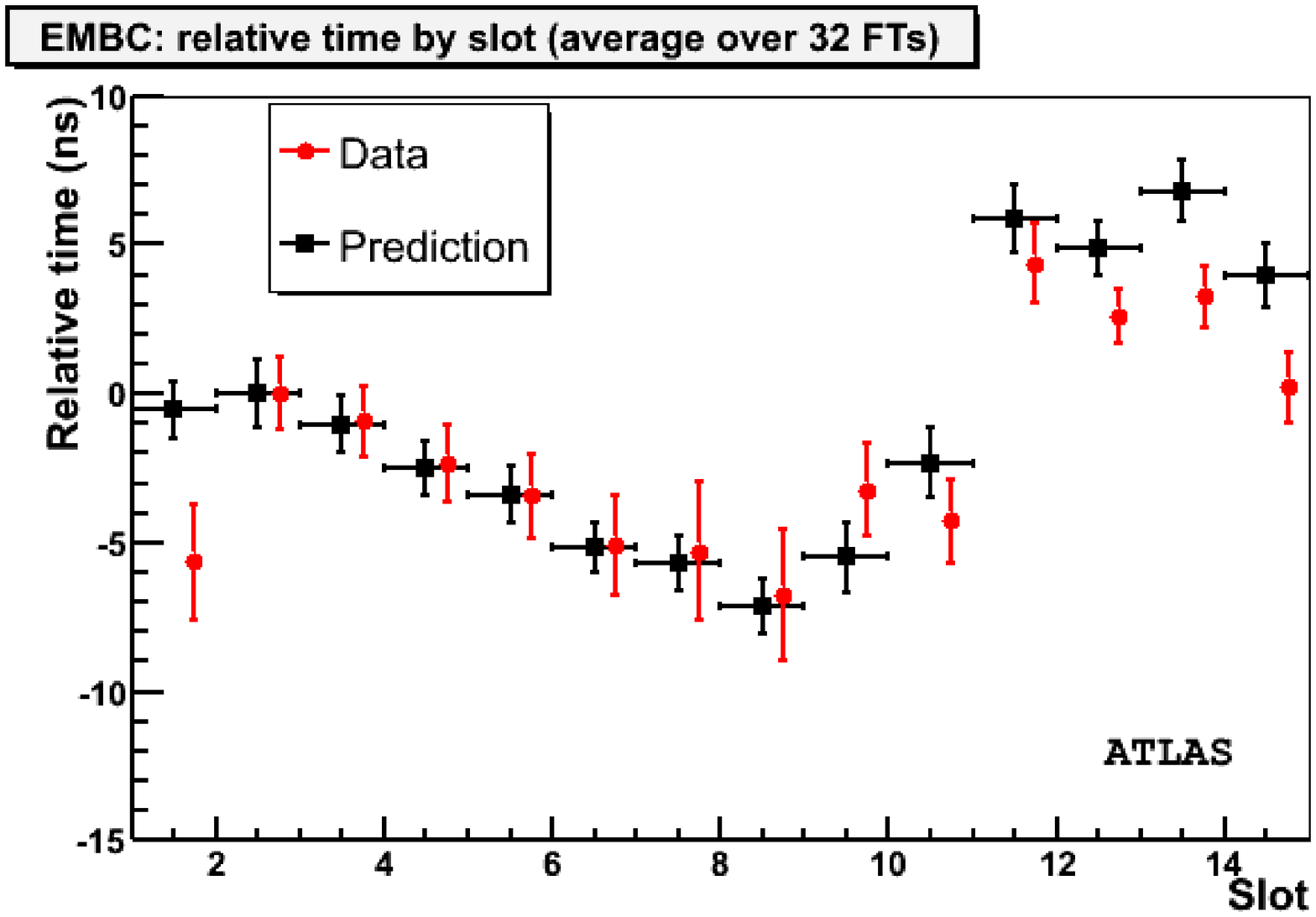}
(b)
\end{minipage}
\caption{ (a) Accumulated energy per layer 2 EM cell in single beam events. (b) Comparison between TOF corrected time in single beam events with the prediction accounting for known delays in readout.}
\end{figure}


\begin{theacknowledgments}
This note reflects the dedicated work of the entire Liquid Argon Calorimeter Group.
\end{theacknowledgments}


\end{document}